\documentclass[12pt]{iopart}

\usepackage{graphicx}

\begin{document}

	\article[Smooth holographic trap for cold atoms]{Central European Workshop on Quantum Optics 2010}{Smooth, holographically generated ring trap for the investigation of superfluidity in ultracold atoms}
	\author{Graham D Bruce, James Mayoh, Giuseppe Smirne, Lara Torralbo-Campo and Donatella Cassettari}
	\address{Scottish Universities Physics Alliance, School of Physics and Astronomy, University of St Andrews, St Andrews, Fife KY16 9SS, Scotland}
	\ead{gdb2@st-andrews.ac.uk}
	\begin{abstract}
		We discuss the suitability of holographically generated optical potentials for the investigation of superfluidity in ultracold atoms.  By using a spatial light modulator and a feedback enabled algorithm we generate a smooth ring with variable bright regions that can be dynamically rotated to stir ultracold atoms and induce superflow.  We also comment on its future integration into a cold atoms experiment.
	\end{abstract}

	\section{Introduction} \label{sect:Intro}
	
	In 1938, Kaptiza \cite{Kapitza_38} and Allen / Misener \cite {Allen_38} independently discovered that, below 2.17~K, liquid Helium-4 can flow without friction.  This superfluidity is also a property of dilute ultracold atomic gases, and has been demonstrated through the presence of a critical velocity \cite{Raman_99}, the formation of arrays of vortices under rotation \cite{Abo-Shaeer_01} and the observation of persistent flow \cite{Ryu_07}.  The high degree of control over both the gas interactions and the trapping potential allows detailed studies of superfluidity in these systems, whilst their accurate modelling is also possible using the mean-field Gross-Pitaevskii equation.
	
	There have been numerous alternative proposals for toroidal atom traps using magnetic fields with \cite{Morizot_06} or without rf-dressing \cite{Gupta_05}, electrostatic fields \cite{Hopkins_04}, time-averaged optical fields \cite{Schnelle_08} or Laguerre-Gauss beams \cite{Wright_00, Franke-Arnold_07}.  An advantage of using optical rather than magnetic potentials is that they are suitable for trapping atoms in any state. In order for any of these to be used to study superflow of ultracold atoms, a stirring mechanism needs to be introduced, such as the transfer of orbital angular momentum from an additional light source \cite{Ryu_07} or a stirring laser beam \cite{Raman_99}. We propose a new method of generating optical ring traps with which to study superfluidity phenomena.  The trapping potential is created by a computer generated hologram (CGH) and a phase-only Spatial Light Modulator (SLM), and incorporates a controllable stirring mechanism using the same laser beam.  
	
	Our SLM is a programmable diffractive optical device which consists of a two-dimensional array of pixels. Each of these is individually controllable, imprinting an arbitrary phase onto incident light.  If the diffracted light is then focussed by a lens, the intensity distribution in the focal plane of the lens is given by the Fourier transform of the phase pattern on the SLM, allowing freedom to create arbitrary intensity distributions. By displaying a sequence of phase patterns on the SLM, dynamic light patterns can also be generated.

	CGHs are becoming increasingly popular as a method to trap ultracold atoms.  SLMs have been proposed as a tool for generating a number of exotic trapping geometries beyond those achievable with standing-wave technologies, such as atom interferometers \cite{McGloin_03} and ring lattices \cite{Franke-Arnold_07}, while they have been used to manipulate single atoms \cite{Bergamini_04}, clouds of cold atoms from a Magneto-Optical Trap \cite{Rhodes_06} and a Bose-Einstein condensate \cite{Boyer_06}.  All works to date have used either arrays of dipole potentials or patterns for which the phase is well known, e.g. Laguerre-Gauss beams.
	
	In general, calculating a phase pattern which will produce an arbitrary desired intensity distribution is not an easy task as the undefined phase in the Fourier plane gives the problem many solutions. However, the calculation can be performed using relatively slow Direct Search Algorithms \cite{Seldowitz_87}, computationally-demanding but highly-accurate Genetic Algorithms\cite{Mitchell_98} or computationally-efficient Iterative Fourier Transform Algorithms such as Gerchberg-Saxton \cite{Gerchberg_72} or Adaptive-Additive \cite{Soifer_97}. In 2008, Pasienski and DeMarco introduced a variant of these algorithms: the Mixed-Region Amplitude Freedom (MRAF) algorithm \cite{Pasienski_08}. This allows the creation of smooth, continuous two-dimensional optical traps.  However, until now experimental achievement of optical traps suitable for ultracold atoms using this algorithm has not been shown.
	
	By recording intensity patterns on a CCD camera, we show that the MRAF algorithm can generate light patterns suitable for trapping a Bose--Einstein condensate.  We further demonstrate that this trap can be dynamically varied in order to induce superflow.  We also find that the physical implementation of these CGHs using SLMs is susceptible to aberrations in the optical system and imperfect device response, thus introducing errors and roughness that were not present during the calculation.  However, we show that this roughness can be reduced by incorporating the MRAF algorithm into a feedback loop.  Finally, we discuss the future integration of our trap into a cold atoms experiment.
	
	\section{Inducing Superflow with a Holographic Optical Trap} \label{sect:Holography}
	
\begin{figure}[htb]
			\begin{center}
					\includegraphics[width=0.5\textwidth]{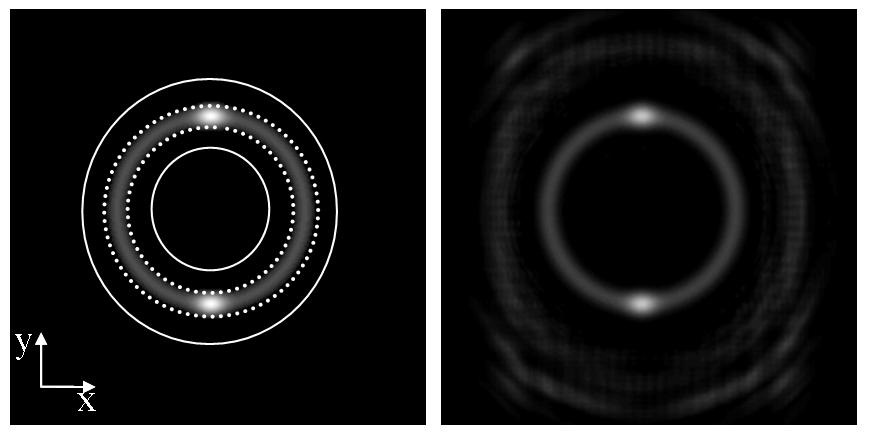}
			\end{center}
  		\caption{(Left) Our target intensity profile showing the measure region (ring enclosed by \dotted) and signal region (ring enclosed by \full).  (Right) The computed output of the MRAF algorithm, showing an intensity which differs from the target by $0.6\%$ in the measure region. This has been achieved by allowing unwanted light to be present outside the signal region.}
				\label{fig:TargPred}
		\end{figure}

	A conservative trapping potential for ultracold atoms can be generated with focussed laser light far-detuned from an atomic transition.  Light of intensity $I(\bf r)$ which is detuned by $\delta$ from an atomic transition frequency $\omega_0$ gives a trapping potential
	
	\begin{equation}
		U({\bf r}) \approx \frac{3 \pi c^2 \Gamma}{2 \omega_{0}^3 \delta} I({\bf r}) \label{eqn:trappingpot}
	\end{equation}
	
\noindent where $\Gamma$ is the natural linewidth of the atomic transition and $c$ is the speed of light.  We design an annular light pattern of intensity $I_0$ with two bright spots of intensity $I_0 + I_p$, given by
	
	\begin{eqnarray}		I(x,y) = & I_0 \exp(-2 (\sqrt{x^2+y^2}-R)^2/w^2) \nonumber \\
 & + I_p \exp(-2 ((x+R \sin \theta)^2+(y-R \cos \theta)^2)/w^2) \\
 & + I_p \exp(-2 ((x-R \sin \theta)^2+(y+R \cos \theta)^2)/w^2) \nonumber 
 \label{eqn:Ir}
	\end{eqnarray}
	
	\noindent where $R$ is the radius of the ring, $\theta$ is the angular position of the bright spots, and $w$ is the $1/e^2$ waist of the Gaussian ring potential, as shown in Figure \ref{fig:TargPred}.  The MRAF algorithm divides the output plane into three regions: the measure region, the signal region and the noise region.  The measure region closely matches the boundaries of the light pattern, whilst the signal region is large enough to contain the measure region plus a border which will be devoid of light.  The amplitude of the light in these regions is fixed to match the target amplitude, meaning ``noise'' on the pattern can only exist in the noise region (the entire plane outside the signal region), where the amplitude is unconstrained.  The algorithm converges to a solution within 100 iterations and the rms error of the calculated output is $0.6\%$.
	
\begin{figure}[htb]
		\begin{center}
			\includegraphics[width=0.75\textwidth]{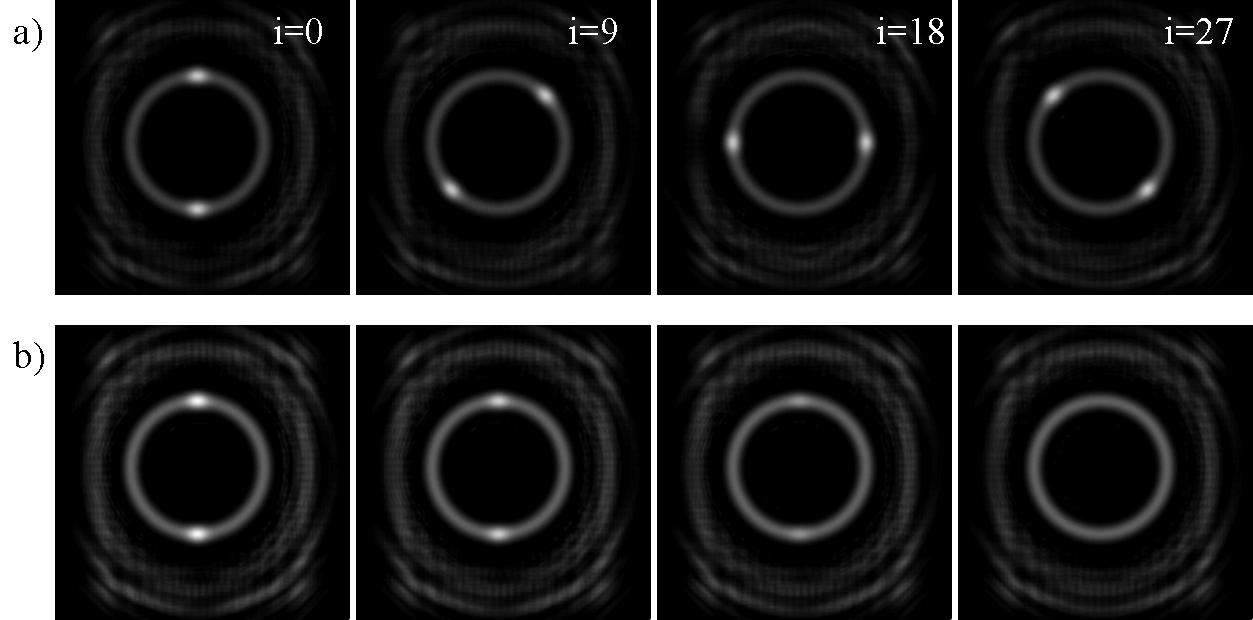}
		\end{center}
		\caption{a) The pattern can be rotated to stir the ultracold gas.  We show equally-spaced snapshots from the first half of a 70-step sequence. b) Once stirring has been completed, the bright spots can be slowly removed to leave a uniform ring trap.}
		\label{fig:rotred}
		\end{figure}

		We plan to induce superflow by ``stirring'' the atoms in the ring potential by rotating the bright spots within the pattern.  This is possible by replacing the phase pattern on the SLM with one which corresponds to the next position of the bright spots, as shown in Figure \ref{fig:rotred}a.  The experiments reported in \cite{Ryu_07} showed superflow of atoms with angular momentum $\hbar$ and $2\hbar$, which for their trap corresponds to a rotational frequency of around $7~$Hz and $14~$Hz respectively.  The maximum rate at which our SLM can refresh the whole pattern is $500~$Hz, which allows a maximum of $70$ steps within one complete rotation of our ring trap.  The difference between consecutive positions of the bright spots should thus be sufficiently small that the motion of the stirring appears continuous to the atoms.  Once superflow has begun, the stirring beams can be gradually removed, as shown in Figure \ref{fig:rotred}b.  As in \cite{Abo-Shaeer_01}, we plan to ramp the value $I_p$ to zero within $20~$ms, which allows us to refresh the pattern ten times.
						
		\section{Experimental Implementation}
		
		Our SLM (Boulder Nonlinear Systems P256) has $256\times256$ nematic liquid crystal pixels, each of which can shift the phase of an incoming laser beam in steps of $2\pi/256$.  We send the phase pattern calculated using MRAF to our SLM, which is illuminated with collimated and spatially filtered light from a $1060~$nm diode laser. To illuminate the SLM as uniformly as possible, we expand the beam to a $1/e^2$ radius of $6~$mm to overfill the active area of the SLM, a square of side length $6.14~$mm.  This ensures that all pixels of the SLM are illuminated by a comparable light power.  MRAF is capable of calculating the phase required to generate the desired intensity distribution regardless of the incoming beam profile, so in future the overfilling could be relaxed to allow more power in the optical trap.  The fact that MRAF shifts light into the noise region coupled to losses due to our overfilling and to the generation of other diffracted orders gives us $3\%$ of the incoming light in the actual optical trap.  However, this low efficiency does not limit the application of the optical traps.
		
		Whilst we have used a limited power for diagnostics, our SLM can be illuminated by up to $3.5~$Wcm$^{-2}$.  We could, for example, illuminate the SLM with $360~$mW of $1060~$nm light to generate a ring of radius $80~\mu$m and waist $25~\mu$m, giving us $10~$mW in the measure region; this corresponds to a ring trap of depth $52~$nK for $^{87}$Rb atoms.  The light pattern shown here is two-dimensional, and disperses quickly away from the Fourier plane.  To confine atoms to the Fourier plane we will add an orthogonal light sheet with a trapping frequency close to that of the ring trap.  The chemical potential of a Bose--Einstein condensate in a ring trap is calculated using the Thomas-Fermi approximation to be
		
		\begin{equation}
			\mu=\hbar \omega \sqrt{\frac{2 N a_s}{\pi R}}
			\label{eqn:mu}
		\end{equation}
		
\noindent where $R$ is the radius of the ring as before, $\omega = 2\pi \times 20~$Hz is the radial trapping frequency, $N$ is the number of trapped atoms and $a_s$ is the s-wave scattering length.  For $2\times10^5$ $^{87}$Rb atoms in the ring trap described above, the chemical potential is $18~$nK.  This is around one third of the trap depth given above, thus making trapping possible.
		
				\begin{figure}[htb]
			\begin{center}
					\includegraphics[width=0.6\textwidth]{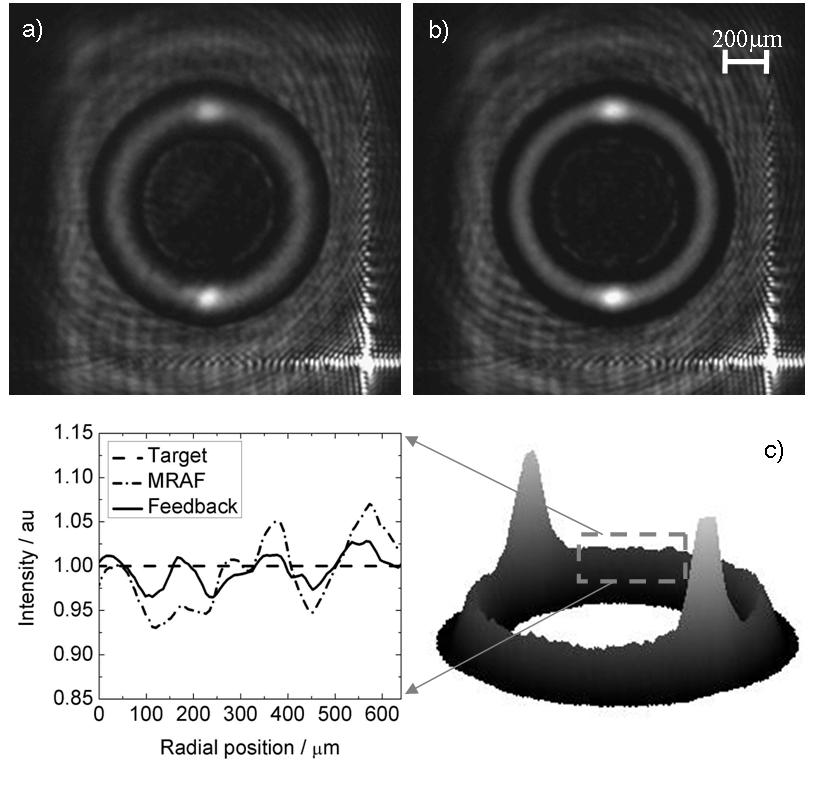}
			\end{center}
  		\caption{a) Initial output of MRAF algorithm. The bright cross at the bottom-right of the image is zeroth order of diffraction by the SLM. b)  Light intensity after seven iterations of the feedback loop. c) The final light intensity within the ring portion of the trap (\full) is significantly smoother and more accurate than the initial output (\chain).  The desired intensity distribution is shown (\dashed) for reference.}
				\label{fig:Outputs}
		\end{figure}		

Whilst the computed patterns give smooth and accurate trapping potentials, imperfect device response and aberrations cause significant discrepancies and roughness in the resulting light pattern shown in Figure \ref{fig:Outputs}a.  We improve the light pattern's suitability as an atom trap by including the algorithm in a feedback loop.  Placing a CCD camera (Thorlabs DCU200 Series) in the Fourier plane of a $f=150$mm lens allows us to measure the difference between the target profile and the real recorded intensity pattern.  We then create a new target pattern by subtracting the discrepancy from the original target, and re-run the MRAF algorithm in an attempt to minimise this difference.  After a few iterations, the measured light profile shown in Figure \ref{fig:Outputs}b is noticeably closer to the original target.  To illustrate this, we plot the intensity of light around the trapping minimum of a section of the ring by integrating over the pixels within a small angle, as shown in Figure \ref{fig:Outputs}c.  
				
		It follows from equation \ref{eqn:trappingpot} that any roughness in the intensity of the light pattern produces a roughness in the energy landscape, which may in turn cause heating or fragmentation of the atom cloud.  Our feedback loop reduce the rms error from $8.48\%$ to $3.88\%$ at the minimum of our optical trap.  This gives us a roughness of $\mu/9$, lower than the value reported in \cite{Ryu_07}, which should allow us to observe superflow in this trap.
		
		As experiments with cold atoms are carried out under vacuum, the windows of the vacuum system will themselves introduce aberrations which would not be accounted for by the feedback loop in its present form.  However, atoms are a very sensitive probe of roughness of the trapping potential, so we envisage continuing to use the feedback loop by taking absorption images of the trapped atoms rather than directly imaging the light profile, and use these images within the feedback loop.

		\section{Conclusion}
		
		We have shown that computer generated holographic techniques supplemented with a feedback algorithm can produce flexible, smooth, all-optical traps for the investigation of superfluidity and persistent currents in ultracold atoms.  The trap depth at different points of the pattern can be dynamically varied and the whole pattern can be rotated to induce superflow.  The geometry created using this method need not be restricted to the one presented here: the creation of ring lattices or ring potentials with areas of lowered intensity for the investigation of Josephson tunnelling, or increasingly non-trivial geometries, should be possible.  For example, by creating two of these traps which intersect and rotate in opposite directions, one could investigate an area of crossing superfluids.  An appropriate choice of trapping light wavelength will allow mixtures of ultracold bosons and fermions and their corresponding superfluidities to be investigated.  Improved understanding of superfluidity in ultracold atoms may help to solve more difficult problems in condensed matter physics, such as the much-sought explanation of high-$T_c$ superconductors.
		
	\ack
	This work is supported by the UK EPSRC.  GS acknowledges support from a Scottish Universities Physics Alliance (SUPA) Advanced Fellowship.  The authors would like to acknowledge useful discussions with Vincent Boyer, Tiffany Harte and Sarah Bromley.
	\bibliographystyle{unsrt}
	\section*{References}

%
\end{document}